\documentclass[useAMS,usenatbib]{mn2e}
\usepackage{psfig}
\usepackage{epsfig}
\bibpunct{(}{)}{;}{a}{}{,}

\def\mnras{MNRAS}
\def\apj{ApJ}
\def\msol{M$_\odot$}
\begin{document}
\title[The innermost stable circular orbit in 4U~1636--536]{Supporting evidence for the signature of the innermost 
stable circular orbit in Rossi X-ray data from 4U~1636--536}
\author[Barret, Olive, Miller]{Didier Barret$^{1}$,
  Jean-Francois Olive$^{1}$, \& M. Coleman Miller$^{2}$
  \thanks{E-mail: Didier.Barret@cesr.fr}\\
$^1$Centre d'Etude Spatiale des Rayonnements, CNRS/UPS, 9 Avenue du
Colonel Roche, 31028 Toulouse Cedex 04, France\\
$^2$Department of Astronomy, University of Maryland, College Park, MD
20742-2421, United States}
\date{Accepted XXX. Received 2006 XXX; in original form XXX }

\maketitle

\begin{abstract}

Analysis of archival Rossi X-ray Timing Explorer (RXTE) data on neutron star low-mass X-ray
binaries has shown that for several sources the quality factor
$Q$ of the lower kilohertz Quasi-Periodic Oscillations (QPO) drops sharply beyond a certain
frequency.  This is one possible signature of the approach to
the general relativistic innermost stable circular orbit
(ISCO), but the implications of such an interpretation for
strong gravity and dense matter are important enough that it
is essential to explore alternate explanations. In this
spirit, M\'endez has recently proposed that $Q$ depends
fundamentally on mass accretion rate (as measured by spectral
hardness) rather than the frequency of the QPO.  Specifically,
he has suggested that analysis of multiple sources shows
trends in $Q$ similar to those previously reported in
individual sources, and he surmises that the ISCO therefore
does not play a role in the observed sharp drop in $Q$ in any
source.  We test this hypothesis for 4U~1636--536 by measuring precisely spectral colors simultaneously with the lower QPO frequency and $Q$ after correction for the frequency drift, over a data set spanning eight years of RXTE observations. We find that in this source there is no correlation between $Q$ and spectral hardness. In particular, no apparent changes in hardness are observed when $Q$ reaches its maximum before dropping off. We perform a similar analysis on 4U~1608--522; another source showing a sharp drop in the quality factor of its lower kHz QPO. We find that for this source, positive and negative correlations are observed between spectral hardness, frequency and $Q$. Consequently, if we are to search for a common explanation for the sharp drop in the quality factor seen in both sources, the spectral hardness is not a good candidate for the independent variable whereas the frequency remains. Therefore, we conclude that the ISCO explanation is viable for 4U~1636--536, and thus possibly for others.
\end{abstract}

\begin{keywords}
Accretion - Accretion disk, stars: neutron, stars: X-rays
\end{keywords}

\section{Introduction}

Kilohertz quasi-periodic brightness oscillations (kHz QPOs) have
been observed with the {\it Rossi} X-ray Timing Explorer  ({\it
RXTE}, Bradt, Swank and Rothschild, 1993) from some 25 neutron star 
low-mass X-ray binary systems
(NS LMXBs).  These QPOs are strong (fractional rms amplitudes
are often $>$10\% in the 2-60~keV band), sharp (quality factors
up to $Q\equiv \nu/{\rm FWHM}\sim 200$), high frequency
(commonly 700--1000~Hz, with the highest claimed detection at
1330~Hz for 4U~0614+091; van Straaten et al. 2000), and
substantially variable (QPO frequencies in a given source can
change by hundreds of Hertz).  They also commonly come in a
pair, with the separation between the upper and lower kHz QPO
staying close to the spin frequency $\nu_{\rm spin}$ or half the
spin frequency despite variations in both QPOs (van der Klis 2006).

There is as yet not a complete consensus about the physical
processes that generate these QPOs, let alone their high quality
factors which pose severe constraints on all existing models (Barret
et al. 2005).  Suggestions include beat-frequency mechanisms
(Miller, Lamb \& Psaltis 1998), some manifestation of geodesic
frequencies (Stella \& Vietri 1999), and resonant interactions
(Abramowicz et al. 2003, Abramowicz et al. 2005).  There is,
however, broad agreement that the upper kHz QPO frequency is close
to the orbital frequency at some special radius (or the vertical
epicyclic frequency in some resonance models, but this is within a
few Hertz of the orbital frequency for the relevant stellar spin
parameters).  By itself, this implies that there must be an upper
limit to the QPO frequency.  For very hard high-density equations of
state or very low-mass stars the limiting frequency could in
principle be set by the orbital frequency at the stellar surface,
but in realistic cases the maximum will instead be determined by the
innermost stable circular orbit (ISCO).  If observational signatures
of the approach to the ISCO were observed, this would confirm the
strong-gravity prediction of unstable orbits, which has no parallel
in Newtonian gravity.  It would also allow us direct measurement of the
mass of the neutron star (see discussion in Miller, Lamb, \& Psaltis
1998), hence the search for such signatures is of great importance
for fundamental physics and astrophysics.

It was proposed theoretically (Miller, Lamb, \& Psaltis 1998, see also Kluzniak, Michelson, \&
Wagoner 1990)
that as the radius that determines the upper kHz QPO frequency
approaches the ISCO (and thus as the lower peak approaches the
ISCO frequency minus $\nu_{\rm spin}$ or $\nu_{\rm spin}/2$),
this will lead to (1)~asymptoting of the frequency to a
limiting value, (2)~decrease in the amplitude of the
oscillation, and (3)~sharp decrease in the quality factor
$Q\equiv \nu/{\rm FWHM}$ of the oscillation.  Zhang et al.
(1998) suggested that the first of these signatures is apparent
in the data from 4U~1820--30, but complications in the relation
between countrate and frequency (M\'endez et al. 1999) have
made the interpretation of this result uncertain (see however,
e.g. Bloser et al. 2000).  More recently, analysis of archival
{\it RXTE} data from multiple sources has revealed a sharp drop
in $Q$ for the lower QPO with increasing  frequency, and this
drop is qualitatively and quantitatively consistent with what
is expected for the approach to the ISCO (Barret, Olive \& Miller 2005a,b
2006).

If confirmed, this result is of great fundamental importance.
It is thus essential to examine alternate explanations.  In
particular, as discussed in Barret, Olive, \& Miller (2006),
there are many factors that collectively determine $Q$ for
an oscillation.  Theoretical arguments (Miller, Lamb, \& Psaltis
1998) as well as recent observational results (Gilfanov \& Revnivtsev 2005)
suggest that although the high observed amplitudes require that
the energy we see in the QPO is liberated at the stellar surface,
the frequency and sharpness of the QPO is determined in the
accretion disk.  In such a picture, in which the frequencies
are generated in some special annulus of the disk, $Q$ depends
on the width of the annulus, the inward radial drift speed,
and the number of cycles a given oscillation lasts.  As discussed
in Barret, Olive, \& Miller (2006), approach to the ISCO can
affect the first two of these factors in a way that agrees with
the data.  

However, it is also possible that other,
non-spacetime-related, effects play a role, e.g., plasma processes 
in the disk, corona, or stellar surface, or interaction with the
stellar magnetic field.  Without a detailed model of this type
one cannot rule definitively for or against such ideas.  Generically,
though, one expects that such factors in $Q$ will depend fundamentally
on the mass accretion rate, whereas the ISCO-related effects depend
fundamentally on the spacetime.  Therefore, a strong correlation
between $Q$ and a proxy for the mass accretion rate would suggest
plasma or magnetic field interactions, whereas the lack of such a
correlation combined with the observed dependence of $Q$ on
frequency would argue in favor of an ISCO interpretation.

Recently, M\'endez (2006) compiled data from multiple sources and
suggested that it is in fact the spectral hardness (his measure
for the mass accretion rate) that is the primary factor in
determining $Q$.  Here we test this suggestion with {\it RXTE}
data on 4U~1636--536.  In \S~2 we discuss our selection of data
and processing algorithms.  We also present our results, and
specifically show that there is no apparent correlation between
the quality factor of the lower kHz QPO and the spectral hardness.  Therefore,
for this source and perhaps others, there is no evidence that the
accretion rate is the primary determinant of $Q$.  We discuss the
implications in \S~3.

\section{Data analysis and results}

We used the data presented in Barret, Olive \& Miller (2005a,b, 2006). The same analysis scheme applies for the data selection. We consider all
data recorded up to September 2004. All PCA Science Event files
were retrieved from the HEASARC archive. A file represents a
temporally contiguous collection of data from a single pointing.
571 files are considered here. They have been filtered for X-ray
bursts and data gaps. Leahy normalized Fourier power density
spectra were computed between 1 and 2048 Hz, over 8 second
intervals with a 1 Hz resolution.

In parallel, we have analyzed the PCA Standard 2 data (a
collection of 129-channel spectra accumulated every 16 seconds),
following standard recipes, using REX 0.3\footnote{http://heasarc.gsfc.nasa.gov/docs/xte/recipes/rex.html}.
We filtered the data using standard criteria: Earth elevation
angle greater than 10 degrees, pointing offset less than 0.02
degrees, time since the peak of the last SAA passage greater than
30 minutes, electron contamination less than 0.1.  The background
of the PCA has been estimated using {\it pcabackest 3.0}, and the
latest bright background model as recommended for sources brighter
than 40 counts/s/PCU. To avoid any possible discontinuity near the loss of its propane layer (in May 2000), we exclude PCU 0 in our analysis. PCU units 2 and 3 provide a good overlap between the Standard 2 and Science Event data (they both provide twice as much data as PCU unit 1 for instance). For each ObsID, for the 2 PCU units, considering only the the top layer (layer 1), we have first generated a response matrix using the latest version
of {\it pcarsp 10.1}. For comparison with previous works, we intend to compute the colors from data
recorded in 4 adjacent energy bands: 3.0-4.5 keV, 4.5-6.4 keV,
6.4-9.7 keV, 9.7-16 keV. For each ObIDs, we read out from the response matrix the relative
channel values corresponding to these energy boundaries, and we
use the {\it FTOOLS} {\it chantrans} to convert the latter into their
absolute channel values, as needed for {\it saextract}, called by
{\it REX}. Since the energy boundaries are not exactly equal to the
above exact values defined above and change in time with the
detector gains, a correction must applied. Following Barret \&
Olive (2002), for each ObsID, we have extracted for each PCU unit
a PHA count spectrum. By fitting the count spectrum with a
polynomial function, one can compute the exact number of counts
within the exact energy bounds. This gives us an average
correction factor,  which corresponds to the difference between
the counts extracted with {\it saextract} and the corresponding counts in the
exact energy bands. Finally, another smaller correction is applied to account for the fact that when the energy boundaries move, the corresponding effective areas also change. The light curves in the four energy bands are normalized to the same effective area, which is, for each ObsID, estimated directly from the response matrix generated for each PCU. After these two corrections, the light curves become all comparable. These
corrections are especially important because the observations span from 1996 to 2004.
\begin{figure} 
\includegraphics[width=0.495\textwidth]{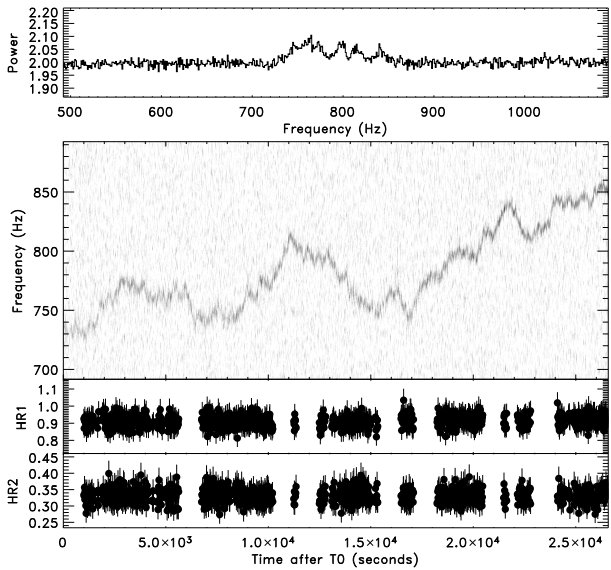}
\caption[]{The Leahy normalized Power Density Spectrum averaged
over the Science Event file is shown at the top.  In the middle
panel, a dynamical PDS is shown. The image has been smoothed to
make the QPO appear more clearly. The bottom two panels
represents the simultaneous evolution of the soft and hard
colors (HR1 and HR2, respectively) as computed from Standard 2
data. The non complete overlap between the two data sets is because the Standard 2 data (only PCU 3 data are
considered) are filtered with more stringent criteria than the
Science Event data. Note that while the QPO frequency varies
from 730 Hz to  850 Hz, there are no appearant changes in
both the soft and hard colors.}
\label{barret_f1}
\end{figure}

We have computed two spectral colors: the soft color (HR1) defined
as the ratio between the 4.5 and 6.4 keV counts and 3.0 and 4.5
keV counts, and the hard color (HR2) as the ratio between the 9.7
and 16.0 keV counts and 6.4 and 9.7 keV counts. For a given
Science Event file, the end-product of our analysis at this first
stage can be summarized in Figure \ref{barret_f1} where the averaged PDS over
the file is shown, together with the dynamical power density
spectrum and the soft and hard colors as derived from the Standard 2
data.

\begin{figure} 
\includegraphics[width=0.495\textwidth]{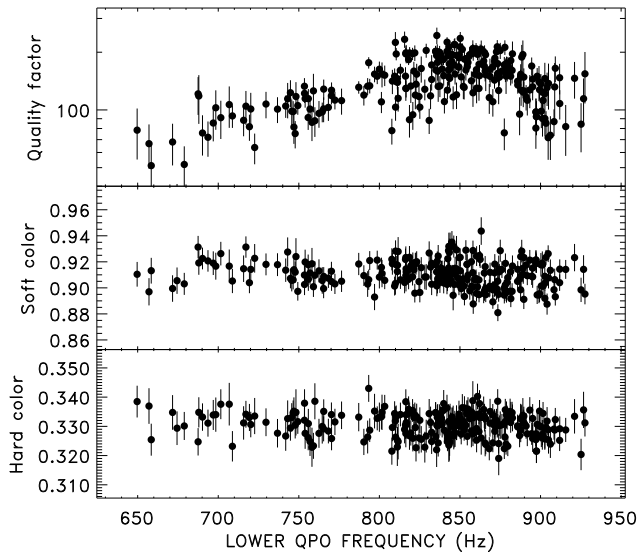}
\caption[]{The quality factor-frequency dependency is shown on the
top panel. Each point is obtained by shifting-and-adding all 8 second PDS
over segments of 1024 second duration (see text). Only points of significance
greater than 4$\sigma$ are shown. The rise of the quality factor with frequency, followed by a sharp drop, is clear, albeit with larger scatter than in Barret et al. (2005a), who averaged the QPO over longer integration times (typical duration of an ObsID, i.e. 4000 seconds). The mid and bottom panels show
the soft and hard colors measured from PCU 3 data as a function of frequency. Only points
for which the overlap between the Science Event data and the
Standard 2 data is larger than 50\% are shown. There is a slight
continuous anticorrelation between soft colors and frequency, but
no correlation at all between frequency and hard color. A fit by
a straight line yields a $\chi^2$ of 230 for 256 degrees of
freedom.}
\label{barret_f2}
\end{figure}

\subsection{Quality factor and spectral colors against frequency in 4U~1636--536} 
For the sake of clarity, our analysis here is focussed on the lower
kilo-Hz QPO, for which we have argued that the drop of coherence at
some critical frequency may be related to an approach of the ISCO
of the region from which the oscillation originates. We wish to
study the dependency of the quality factor versus the soft and hard
colors. One would expect that if the drop is indeed related to a
spacetime effect, it is not primarily dependent on the energy
spectrum. This idea can be tested with spectral
colors, which allow us to search for subtle spectral variations.

In order to estimate the quality factor of the QPOs, one must first
correct for the frequency drifts. Here we use a sliding window
based technique. Namely, we group as many consecutive 8 second PDS
($N\le64$) as needed to detect a QPO with a significance above a
threshold of $3\sigma$. The maximum integration time is
then $64\times8=512$ seconds. In case of no detection within such
an interval starting at $T_0$, a new search starts at $T_0+32$ sec.
The QPO frequency in each 8 second PDS is then estimated using a
linear interpolation between all detected QPO frequencies within a
continuous segment.  We identify those files containing a lower
kilo-Hz QPO in the quality factor-frequency plane (see Barret et
al. 2005a for details). We keep those files in which the quality
factor recovered after correction for the frequency drift is larger
than 30, corresponding to a mean QPO frequency larger than 650 and smaller than
950 Hz.   We then divide the data into segments of 1024 seconds,
and shift-and-add all 8 second PDS within each segments to a
reference frequency which is set to the mean QPO frequency
assigned after interpolation to the 8 second PDS over the continuous segment. Some segments are
not complete and we have removed all those in which the total
number of 8 second PDS shifted is less than 64 (i.e. 512 seconds).
Within each segment, when there is an overlap of at least 50\% with
the Standard 2 data, we compute the corresponding colors HR1 and
HR2.  In Figure \ref{barret_f2}, we plot the resulting quality factor and
spectral colors. No dramatic changes in either the soft and hard colors
is observed when the quality factor of the lower kilo-Hz QPO shows
a clear drop. In particular, over the frequency range spanned by
the lower kilo-Hz QPO, the hard color is consistent with being
constant (a fit by a straight line yields a $\chi^2$ of 295 and 230
for 260 and 256 degrees of freedom  (d.o.f.) for PCU2 and PCU 3 data). Over a narrower frequency range (780-930 Hz), Di Salvo et al. (2004) obtained very similar results. There
is a negative smooth anticorrelation between soft color and lower
QPO frequency, but nothing notable around the frequency where the
quality factor of the QPO drops off, around 850 Hz.

\begin{figure}
\centerline{\includegraphics[width=0.45\textwidth]{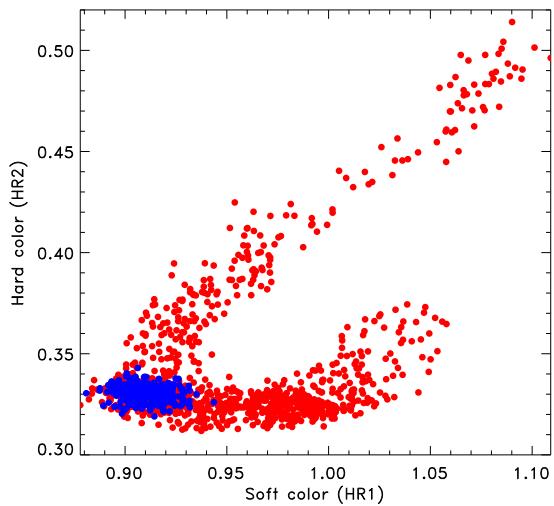}}
\caption[]{The color-color diagram of all the archival RXTE
data analyzed for 4U~1636--536. The color is averaged over
segments of 1024 second duration (red filled circles).
Bursts and gaps have been filtered out. In blue, the colors
corresponding to the QPO detections is shown. Only data
from PCU 3 are shown, but the same results is obtained with
PCU 2 data. }
\label{barret_f3}
\end{figure}

In Figure \ref{barret_f3}, we show the color-color diagram for all the data
considered and overlay the region over which the lower QPO
was studied with the present technique. The
lower QPO is detected only over a very delimited region of
the color-color diagram, even though that it samples a
relatively wide range of frequency and quality factor. This by itself shows that even if subtle correlations are masked by the statistics (but see Fig. 4), the drop of the quality factor is not associated with a dramatic spectral change in this source. It is interesting to note that with the procedure described in this paper, lower kilo-Hz QPOs are detected at the intersection between the bottom and diagonal branches of the color-color diagram.
The count rate varies from 130 counts/s to 340 counts/s in PCU3.

\begin{figure} 
\centerline{\includegraphics[width=0.45\textwidth]{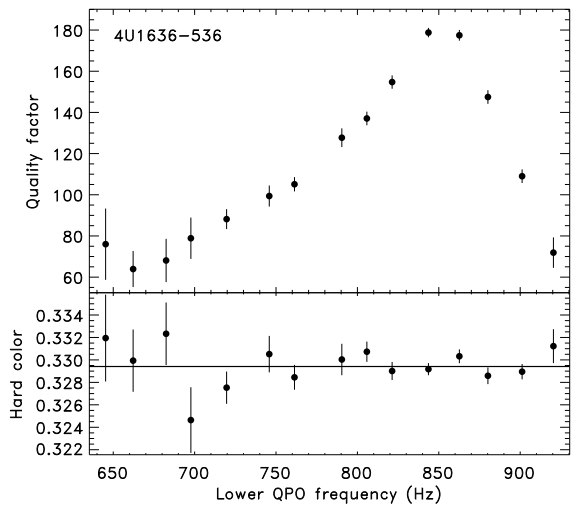}}
\caption[]{The variation of the quality factor and hard color with frequency, grouping the data of Figure \ref{barret_f2} with a frequency bin of 20 Hz. The hard color is consistent with remaining constant over the frequency range sampled by our analysis (a fit yields $\chi^2$ of 15.2 for 14 d.o.f.).}
\label{barret_f4}
\end{figure}

Next, we have grouped the data  of Figure \ref{barret_f2} using weighted averages (and the $1\sigma$ errors) in the frequency space, with a bin of 20 Hz to get sufficient statistics and still enough points to keep a good description of the overall behavior of the quality fact and hard color with frequency. The results are shown in Figure \ref{barret_f4}. The size of the error bars on hard colors has been decreased on average by a factor of 2.7 compared to the data of Figure \ref{barret_f2}; the mean error on the hard color is reduced to 0.0016. A fit with a constant yields a $\chi^2$ of 15.2 for 14 d.o.f. for PCU 2 data (15.5 for 13 d.o.f for PCU 3 data). 

\begin{figure} 
\centerline{\includegraphics[width=0.45\textwidth]{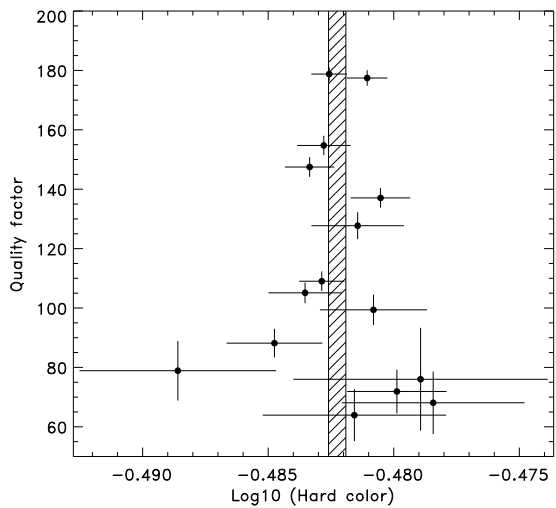}}
\caption[]{Quality factor versus hard color for 4U~1636--536 using grouped data of Figure \ref{barret_f2}. The $1\sigma$ error on the mean hard color is represented by the hashed region. }
\label{barret_f5}
\end{figure}

In Figure \ref{barret_f5}, we plot the quality factor against the logarithm of the hard color to enable a comparison with the data presented for 4U~1608--522 in Mendez (2006). 

\subsection{Comparison with 4U~1608--522}
\begin{figure} 
\includegraphics[width=0.45\textwidth]{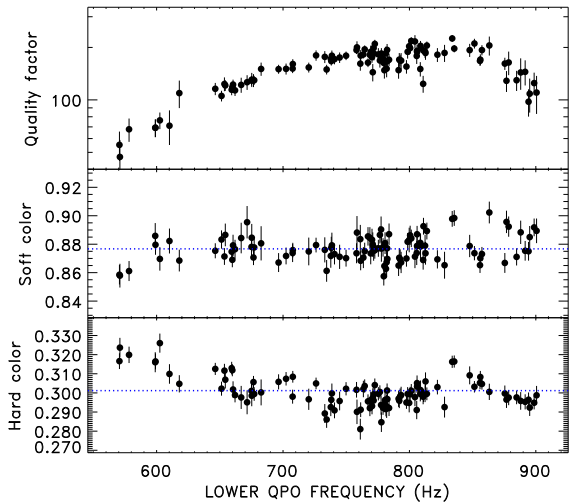}
\caption[]{Same as fig. \ref{barret_f2} but for 4U~1608--522. The quality factor-frequency dependency is shown on the
top panel. Each point is obtained by shifting-and-adding 8 second
over segments of 768 second duration (chosen such that the mean error on the hard color is the same as the one measured for 4U~1636--536 in Fig. \ref{barret_f2}). Only points of significance
greater than 4$\sigma$ are shown. The mid and bottom panels show
the soft and hard colors measured from PCU 2 data as a function of frequency. Only points
for which the overlap between the Science Event data and the
Standard 2 data is larger than 50\% are shown. The same trends are seen in PCU 1 and 3.}
\label{barret_f6}
\end{figure}

Using the same technique as described above, we have reprocessed the data of 4U~1608--522 presented in Barret, Olive \& Miller (2006) focussing on the lower kilo-Hz QPOs. This enables a direct comparison of two homogenous data sets obtained through the same processing scheme. The data set used includes data presented in Mendez et al. (1999), as well as observations performed in september 2002, march 2003 and 2004 (proposal numbers 70059, 80406, 90408). As for 4U1636-536, we identify segments containing a lower kilo-Hz QPO in the quality factor-frequency plane. For 4U~1608--522, as shown in Barret, Olive \& Miller (2006), the lower QPO so identified spans a frequency range going from $\sim 570$ to $\sim 900$ Hz.  PCU 2 provides the best overlap between the Science Event and the standard 2 data for this source. In Fig. \ref{barret_f6}, we show the quality factor, soft and hard colors against frequency. Due to the larger count rate of 4U~1608--522 (varying from 150 counts/s up to 480 counts/s in PCU 2), in order to get similar errors on the hard color as the one of 4U~1636--536, one can use shorter integration time for estimating the QPO parameters. We have used $8\times 96=768$ seconds for 4U~1608--522, instead of 1024 seconds for 4U~1636--536. For 4U~1608--522, the hard color is clearly not consistent with being constant: a fit by a constant yields a $\chi^2$ of  371 for 103 d.o.f. Negative and positive correlations between hard color and frequency are observed (the same behavior is seen in other PCA units). 

\begin{figure} 
\includegraphics[width=0.45\textwidth]{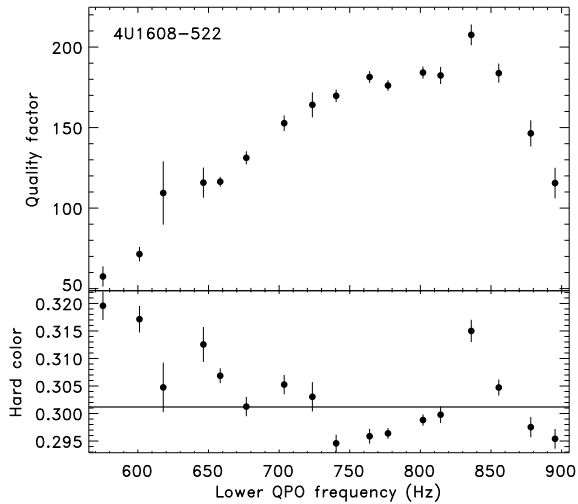}
\caption[]{Same as figure \ref{barret_f4} but for 4U~1608--522. The data are those recorded with PCU 2. Clearly, thanks to the improved statistics, the relationship between the hard color and frequency is revealed to be more complex than previously thought for this source (e.g. Mendez et al. 1999).}
\label{barret_f7}
\end{figure}

\begin{figure} 
\includegraphics[width=0.45\textwidth]{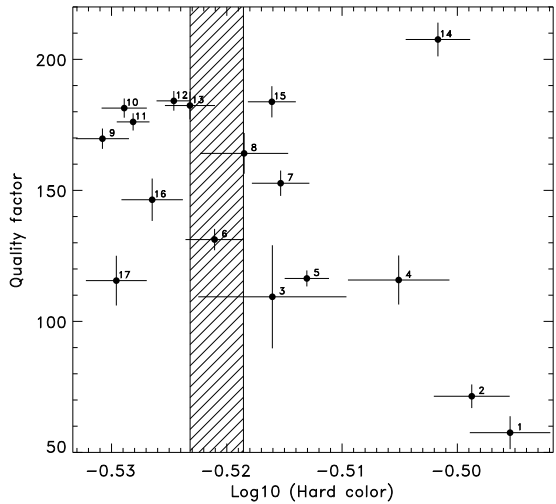}
\caption[]{Same as figure \ref{barret_f5} but for 4U~1608--522. The data are those recorded with PCU 2. The data points are labeled in ascending order with increasing frequency.}
\label{barret_f8}
\end{figure}

We then grouped the data shown on Fig. \ref{barret_f6} also with a bin of 20 Hz to produce Fig. \ref{barret_f7} and \ref{barret_f8}. The mean error on hard colors for 4U~1608--522 is reduced to 0.0020, i.e. slightly larger than those of fig. \ref{barret_f4}. Fitting the hard color with a constant results in a $\chi^2$ of 263 for 16 d.o.f. Restricting the frequency range to above 650 Hz (the minimum frequency in 4U~1636--536), yields a $\chi^2$ of 147 for 13 d.o.f. Since we are interested in the region where $Q$ reaches its maximum before dropping off, one can consider only those points from, say, 100 Hz before the peak (around 840 Hz) up to the last point, at $\sim 930$ Hz in 4U~1636--536 and $\sim 900$ Hz in 4U~1608--522, respectively. We obtain a $\chi^2$ of 9 for 8 d.o.f and 95 for 7 d.o.f for 4U~1636--536 and  4U~1608--522. Clearly for the latter, the hard color is not consistent with being constant over the frequency range spanned by the lower QPO, including the interesting part which encompasses the Q drop-off. We have verified that the same conclusion is reached, independently of the integration time used to estimate the QPO parameters. For instance, using 64 seconds for 4U~1608--522 as in Mendez et al. (1999), the mean error on the hard color after binning is 0.0025 (20\% larger than with 768 seconds), and the $\chi^2$ is 164 for 15 d.o.f. Looking at Fig. \ref{barret_f6}, it is worth noting that the relationship between the hard color and frequency is revealed to be more complex than, and certainly not as smooth as, previously thought, based on measurements obtained with lower statistics (e.g. Mendez et al. (1999)). 

Comparing Fig \ref{barret_f4} and Fig. \ref{barret_f6} requires some comments. First, given that the size of the error bars are fully comparable, it shows that if a similar trend as seen for 4U~1608--522 were to be present in 4U~1636--536, we would have observed it. Second, we note that in 4U~1608--522, significant variations of the hard color (still limited at the $\sim 5$ \% level) are observed around the peak of the quality factor-frequency curve. These variations may reflect some changes in the source behavior (e.g. accretion rate, Mendez 2006), but Fig. \ref{barret_f4} shows that  the effect produced is by no means universal, as we failed to detect it in a similar way in 4U~1636--536. It could be that the drop of the quality factor and change in hard color are simply concomitant in 4U~1608--522. 

From this comparison, one can therefore conclude that if we are to search for a common explanation for the sharp drop in the quality factor seen in both sources, the hard color is not a good candidate for the independent variable whereas the frequency remains. 

\section{Discussion}

The results presented in this paper demonstrate
that the drop of the quality factor of the lower kilo-Hz QPO in
4U~1636--536 is not accompanied by a significant change in the energy
spectrum of the source, as measured by the spectral colors.
Therefore, in this source, if the quality factor depends
fundamentally on mass accretion rate, it must somehow do so in a
way that leaves no detectable correlation between quality factor and either
spectral measures or countrate (see Barret, Olive, \& Miller
2005a,b, 2006).  This seems difficult and contrived.  In contrast,
the strong correlation of $Q$ with frequency in the lower peak is
as expected if it is largely driven by approach to the ISCO.
Together with the observed steady drop in rms amplitude,
saturation of the QPO frequency with increasing count rate, and
the quantitative consistency of ISCO models with the $Q$ versus
frequency curve (Barret, Olive, \& Miller 2005a,b, 2006),
4U~1636--536 behaves as expected if the phenomenology is
linked to the spacetime and not to the mass accretion rate. Other
sources need to be analyzed similarly, but the ISCO hypothesis and
its attendant implications for strong gravity and dense matter are
still entirely viable.

What, then, could be the explanation for the results of M\'endez
(2006), in which he found a correlation between hardness (or average
luminosity) and maximum reported $Q$ over sources spanning a range
of two orders of magnitude in luminosity? The left panel of his
Figure~3 shows that the maximum reported quality factor is low for
the lowest-luminosity source  (4U~0614+091, at $L/L_{\rm Edd}\approx
6\times 10^{-3}$ and $Q_{\rm max}\sim 30$ [Barret, Olive, \& Miller
2006 obtained $Q_{\rm max}\sim 50$ for this source]), high  ($Q_{\rm
max}\approx 100-200$) for sources with $L/L_{\rm Edd}\sim 0.02-0.2$,
and low ($Q_{\rm max}\sim 10-20$) for sources with $L\sim L_{\rm
Edd}$.  A similar pattern, although less monotonic, is shown in his
Figure~4, of $Q$ versus hard color. Because this pattern with
luminosity or hard color is  similar to the behavior in individual
sources with frequency ($Q$ is low at low frequency, rises to a
peak, then drops sharply), M\'endez concludes that it is unlikely
that the ISCO plays a role in any of these systems.

We believe that there is another interpretation.  At the low
luminosity end of M\'endez's correlation there is a single key
source: 4U~0614+091.  Figure~1 of Barret, Olive, \& Miller (2006)
shows that there is no apparent drop in the quality factor of the
lower QPO up to $\sim 700$~Hz. On the other hand,  lower QPOs at frequencies above 700 Hz have been reported  with low Q values (van Straaten et al. 2000), yet without correction for the frequency drift.  The status of 4U~0614+091 is thus different than the status of sources such as 4U~1636--536  for which, thanks to the necessary frequency drift correction applied, a clear maximum has been observed. This is an issue because if the 1330 Hz detection of the upper QPO is real, one would not expect, for any plausible spin frequency of the neutron star, that the frequency at which $Q$ starts decreasing to be $\sim 700$ Hz or so. A careful re-examination of the 4U~0614+091 data is thus underway.

At the high luminosity end, the sources all have
very high luminosity indeed, comparable to Eddington.  Standard
disk accretion theory (e.g., Shakura \& Sunyaev 1973) then suggests
that the disk thickness will be comparable to the orbital radius,
and that as a consequence the inward radial drift speed (which 
scales as $(h/r)^2$, where $h$ is the disk half-thickness) will be
large as well.  As discussed in Barret, Olive, \& Miller (2006),
a large inward speed will necessarily decrease $Q$ regardless of
other factors.  We note that this effect
can also be important in reducing the maximum $Q$ from
$\sim 200$ to $\sim 100$ around a luminosity $L\sim 0.1-0.2L_{\rm Edd}$,
as seen by M\'endez (2006).  It is therefore not surprising that high luminosity
sources have low $Q$, but it is also not relevant to the evaluation
of the behavior of $Q$ with frequency in much lower luminosity
sources.

As a final remark, as we have discussed previously (e.g. Barret,
Olive, \& Miller 2005a,b, 2006), if the ISCO interpretation is
correct for our data, then one can infer a mass of the order of 2
\msol~for the neutron star. This is consistent with
phase-resolved spectroscopy of 4U~1636--536 at the VLT by Casares
et al. (2006), who assume plausible binary parameters
(inclination, disk flaring angle, mass of the donor star) and
infer a mass $1.6-1.9$ \msol~for the neutron star.  In addition,
a variety of modern models for neutron star matter, involving
hyperons, quarks or normal matter,  predict maximum neutron star
masses as large as $\sim 2$\msol~(Jha et al. 2006, Klahn et al.
2006).  Our results thus add to the growing evidence for heavy
neutron stars in accreting systems.

\section{Conclusions}
The case for the ISCO is still promising.  Analysis
of the type that we perform in this paper will be needed for other
sources, to determine the strength of evidence in those cases.
In addition, focused observations or re-analysis of specific objects will be useful, starting with 4U~0614+091.
\section{Acknowledgements}
We thank Mariano M\'endez for stimulating discussions and for helpful interaction along the revision of the paper.  We are also grateful to Jean-Luc Atteia for discussions about statistics related issues, and to Diego Altamirano for initial discussions. MCM
was supported in part by a senior NPP fellowship at Goddard Space
Flight Center.  This research has made use of data obtained from the
High Energy Astrophysics Science Archive Research Center (HEASARC),
provided by NASA's Goddard Space Flight Center.

We thank an anonymous referee for comments that helped to improve the content of this paper.

\end{document}